\documentclass[useAMS,usenatbib]{mn2e}
\usepackage{graphicx}
\usepackage{hyperref}

\title[$k$-d~match]{$k$-d~Match: A Fast Matching Algorithm for Sheared
  Stellar Samples}
\author[J. S. Heyl]{Jeremy S. Heyl$^{1}$\thanks{Email:
    heyl@phas.ubc.ca; Canada Research Chair} \\
$^{1}$Department of Physics and Astronomy, University of British
Columbia, 6224 Agricultural Road, Vancouver, BC V6T 1Z1, Canada}

\begin{document}
\date{Accepted 2013 May 1. Received 2013 May 1; in original form 2013
  April 3}

\pagerange{\pageref{firstpage}--\pageref{lastpage}} \pubyear{2013}

\maketitle

\label{firstpage}

\begin{abstract}
  This paper presents new and efficient algorithms for matching
  stellar catalogues where the transformation between the coordinate
  systems of the two catalagoues is unknown and may include shearing.
  Finding a given object whether a star or asterism from the first
  catalogue in the second is logarithmic in time rather than
  polynomial, yielding a dramatic speed up relative to a naive
  implementation.  Both acceleration of the matching algorithm and the
  ability to solve for arbitrary affine transformations not only will
  allow the registration of stellar catalogues and images that are
  now impossible to use but also will find applications in machine
  vision and other imaging applications.
\end{abstract}


\section{Introduction}

Finding the correspondances between catalogues of objects has long
been a goal of astronomy.  Generally, one first must place both
catalgoues on the same coordinate system and then search in two
dimensions for the nearest neighbour in the second catalogue of each
object in the first catalogue.  Often one does not know the coordinate
transformation between the catalogues initially, so part of the first
step is to determine this transformation.  Finding the transformation
between the coordinate systems also basicially involves a search for
objects that correspond to each other in each catalgoue and are
invariant under the transformation.  \citet{1986AJ.....91.1244G}
introduced the technique of looking for similar triangles in two
catalogues.  The property of similarity is invariant under
translation, rotation, magnification and inversion.  The algorithm
outlined in this paper uses the ratio of sides as the invariant under
the coordinate transformation as in \citet{1995PASP..107.1119V} and
searches for several triangles with similar transformations.  In the
end even a small catalogue can result in a large number of triangles
to search in a catalogue: $N(N-1)(N-2)/6$ or on the order of $N_1^3
N_2^3$ comparisions to make.  Several authors
\citep[e.g.][]{1995PASP..107.1119V} have outlined techniques to
accelerate the calculation of the triangles (${\cal O}(N^2$) vs.\
${\cal O}(N^3)$) at the expense of storing information and
accelerating the search process which decreases the prefactor on
${\cal O}(N_1^3 N_2^3)$ by presorting the triangles
\citep{1995PASP..107.1119V}, weighting the triangles by the magnitudes
of the stars \citep{1994ApOpt..33.4459S}, culling the triangles to
compare \citep{2006PASP..118.1474P} or quitting after only a fraction
of the triangles have been compared and a sufficiently good fit is
determined \citep{2007PASA...24..189T}.

Both finding the matching triangles and later the matching objects in
the catalogues require finding neighbours in a two-dimensional
space.  If the space were one dimensional, one would use a binary tree
to search in $\log N$ time.  \citet{Bentley:1975:MBS:361002.361007}
developed a generalisation of the binary tree for arbitrary number of
dimensions, the $k$-d~tree.  In this case $k$ equals two.  This
algorithm dramatically speeds the search over the two dimensions, and
runs the search for the {\tt astrometry.net} algorithm
\citep{2010AJ....139.1782L}.  In particular, the process of finding
the nearest neighbour in the two catalogues is sped up from ${\cal
  O}(N_1 N_2)$ to ${\cal O}\left [(N_1+N_2)\log N_2\right ]$.  For the
triangle search the improvement is even more dramatic from ${\cal
  O}(N_1^3 N_2^3)$ to ${\cal O}\left [(N_1^3+N_2^3)\log N_2\right ]$.

The dramatic acceleration of the determination of the transformation
encourages a generalisation of the triangle matching technique of
\citet{1986AJ.....91.1244G}.  In particular the property of similarity
of triangles is invariant under translation, rotation, magnification
and inversion but not shearing.  If there is an significant shearing
between the two coordinate systems, a triangle matching algorithm will
fail.  However, it is straightforward to generalise this techinque for
a general affine transformation.  Such a transformation will preserve
the ratio of areas, so the new technique is to build quadrilaterals
from sets of four objects in each catalogue and calculate the ratio of
areas of the triangles that comprise the quadrilaterals.  Only two of
the three area ratios are independent so the final search is again two
dimensional.  However, the number of quadrilaterals is huge
$N(N-1)(N-2)(N-3)/24$ or on the order of $N_1^4 N_2^4$ direct
comparisons to make.  The $k$-d tree accelerates this quadrilateral
search dramatically to ${\cal O}\left [(N_1^4+N_2^4)\log N_2\right ]$,
so it is even faster than the customary direct search over triangles.

The following sections introduce the $k$-d tree data structure
(\S~\ref{sec:k-d-tree}), the quadrilateral search
(\S~\ref{sec:quadrilateral-search}), the elimination of false
positives (\S~\ref{sec:false-positives}), present results of these new
techniques (\S~\ref{sec:results}) and discuss future applications
(\S~\ref{sec:discussion}).

\section{The \lowercase{$k$}-d Tree}
\label{sec:k-d-tree}

\citet{Bentley:1975:MBS:361002.361007} introduced the $k$-d tree as a
multidimensional generalisation of a binary tree.  In a binary tree
each node contains its data, a key and two pointers one toward the
daughter with a larger value of the key and one toward the daughter
with a smaller value of the key.  In the $k$-d tree, the key is
multidimensional.  In the current case the key is two-dimensional.
Again each node contains its data, its {\em two} keys and two pointers one
toward the daughter with a larger value of the first key and one
toward the daughter with a smaller value of the first key.  Looking at
one of the daughter nodes, it contains pointers toward its daughter
with a larger value of the {\em second} key and one toward the
daughter with a smaller value of the {\em second} key.  In this way
each node divides its portion of its space in two with alternatively
horizontal and vectical lines.  The $k$-d tree is straightforward to
generalise to more dimensions.  \citet{Bentley:1975:MBS:361002.361007}
showed the construction and the optimization of the tree requires on
order of $N \log_2 N$ operations and searching the tree requires on
order of $\log_2 N$ operations where $N$ is the number of nodes in the
tree.

A possible alternative data structure is a quadtree
\citep{quadtree} in which each node has four children splitting the
node's space into four quadrants.  This has the disadvantage of that
many more of the pointers within the data structure will be null than
with a $k$-d tree.  Furthermore, although the quadtree may be
generalised to higher dimensions, to implement it efficiently in
various numbers of dimensions may require substantially different
algorithms.  On the other hand, in a $k$-d tree one can alternate over
however many dimensions that the key requires; the false-positive
search (\S~\ref{sec:false-positives}) uses a three-dimensional tree.
\citet{2010AJ....139.1782L} use a $k$-d tree to store four-dimensional
keys representing the shape of four-star asterisms throughout the sky.
They resort to a larger asterism because over the entire sky there
would be many near misses for triangular asterisms given the typical
positional errors in astronomy.

\citet{LangPhD} has developed a very memory efficient $k$-d tree
algorithm for use with {\tt astrometry.net}, and it is publicly
available.  This implementation does not use pointers between the
nodes, but rather stores the nodes in an array such that there is a
straightforward one-to-one correspondance between the position in the
array and the location in the tree structure.  However, matching small
fields will be limited to hundred of stars and possibly millions of
triangles, so memory efficiency is not crucial. The algorithm
presented here uses the implementation of
\citet{tsiombikas11:_kdtree}, available on Google Code.  For larger
datasets, a more memory efficient implementation may be helpful and
could easily replace the library used here.

\section{The Quadrilateral Search}
\label{sec:quadrilateral-search}

\begin{figure}
\includegraphics[width=0.45\linewidth]{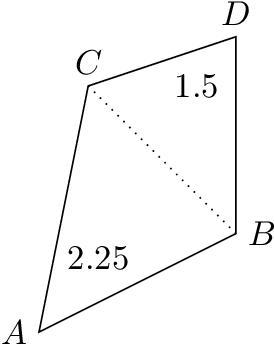}
\includegraphics[width=0.45\linewidth]{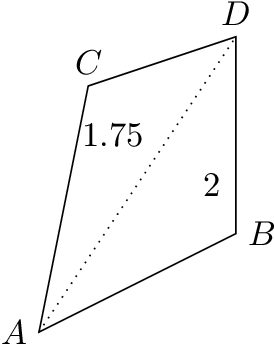}
\caption{A quadrilateral divided into two triangles in two different
  ways by the dotted lines.  On the left are triangles $A$ and $D$,
  and on the right are $B$ and $C$. The areas of each triangle are
  given. The vertices are labeled in order of the area of the triangle
  to which they are adjacent.}
\label{fig:quad}
\end{figure}

The efficient $k$-d tree data structure begs for more complicated
problems to solve.  In particular \citet{2010AJ....139.1782L} decided
to use four-star asterisms to reduce the number of false positives in
their search throughout the sky to find a particular star field.  They
developed a four-component key for each quadrilateral that is
invariant under translations, rotations, scalings and inversion.  For
a triangle, one can only obtain a two-component key, for example the
ratio of each of the smaller sides to the longest side. This
improvement is simple to visualise when one considers a quadrilateral
as the union of two triangles.  Here the challenge will not be to find
a particular field somewhere in the sky, but to match a field with a
given catalogue where the transformation between the coordinate
systems may include shearing as well, i.e. to find the general affine
transformation that relates the two coordinate systems.  A general
affine transformation does not preserve angles, the relative length of
sides or the geometric hash devised by \citet{2010AJ....139.1782L}.
However, since the affine transformation applies the same Jacobian
throughout the field, the ratio of areas of shapes will be preseved.
In particular the ratio of the areas of the various triangles that
comprise a quadrilateral will be preserved.  Given the coordinates of
each of the vertices of the quadrilateral it is straightforward to
calculate the area of the triangle that spans three vertices by a
cross product.  Because one can chose to leave out a vertex from each
triangle, each quadrilateral will generate four triangles as shown in
Fig.~\ref{fig:quad}.  By sorting these triangles in area, the area of
each triangle is normalized by that of the largest triangle.  This
would appear at first to yield three ratios $B/A, C/A$ and $D/A$ where
$A, B, C$ and $D$ are the areas of the triangles in decreasing order.
However, the sum of the areas $B+C$ equals $A+D$ (both pairs make up
the entire quadrilateral), so the ratio $D/A$ is not independent and
equals $D/A=B/A+C/A-1$, so only the two areas $B/A$ and $C/A$ are used
to represent the quadrilateral.

\section{False Positives}
\label{sec:false-positives}

Even a small catalogue will generate a large number of asterisms
(triangles and quadrilaterals), and there are likely to be many near
misses due to coincidences and observational errors.  It is
straightforward to query the $k$-d tree to get all the asterisms whose
keys lie within a sphere surrounding a particular value.  Because the
keys used in this algorithm vary between zero and one, it is
natural to select a tolerance appropriate for the errors in
one's data ($10^{-5}$ for the triangle search and $3 \times 10^{-3}$
for the quadrilateral search).  Larger tolerances will yield more
matching asterisms, so eliminating false positives is crucial. To
identify the best matching asterisms, the transformation between the
coordinate systems defined by the matching asterism is determined.  It
has the following form
\begin{equation}
x_2 = a x_1 + b y_1 + c,
y_2 = d x_1 + e y_1 + f.
\end{equation}
If no shearing is present, $a=e$ and $b=-d$, so the values of $a, b,
c$ and $f$ can uniquely determine the transformation.  The parameters
of the transformation are determined by least-squares fitting over the
points of the matching triangles or quadrilaterals and can include
shear. The values of $a$, $b$, $d$ and $e$ are typically of order
unity while $c$ and $f$ account for translations and may be large;
therefore, for each matching asterism, the points that comprise the
asterism in each catalogue are stored in a second $k$-d tree with the
values of $a$, $b$ and $c/1000$ serving as keys.  The value of $c$
must be scaled by a typical value that one expects for the
translation, so that the three values within the $k$-d tree are of
similar order.

When a new matching asterism is found, its values of $a$, $b$ and $c$
are compared with those of the previous matches.  If a previous match
or matches are found with similar values of $a$, $b$ and $c/1000$
(within a distance of $10^{-3}$ in the space of $a$, $b$, $c/1000$),
the coordinate transformation for all the points in all the matching
asterisms with similar values of $a$,$b$ and $c$ is calculated.  If
the values of $a$, $b$ and $c$ for this transformation are similar to
those for the individual asterisms, it is likely that these are true
matches.  No effort is made to cull repeated stars from the ensemble
of matching asterisms; therefore, stars that are members of several
asterisms have a larger weight in the fit for the transformation.
These repeated stars do not appear multiple times in the final matched star
lists. Because this false positive detection scheme also relies on a
$k$-d~tree but with fewer entries (only the matching asterisms),
it adds little overhead and increases the confidence and the accuracy
in determining the transformation between the catalogues.  The factor
by which the $x-$translation parameter $c$ is scaled can be given by
the user.  Furthermore, if a single transformation can account for at
least 20 asterisms, the algorithm stops and outputs this best fitting
transformation.  Again the number of asterisms to count before
quitting can be changed by the user.

\section{Results}
\label{sec:results}

\begin{figure}
\includegraphics[width=\linewidth]{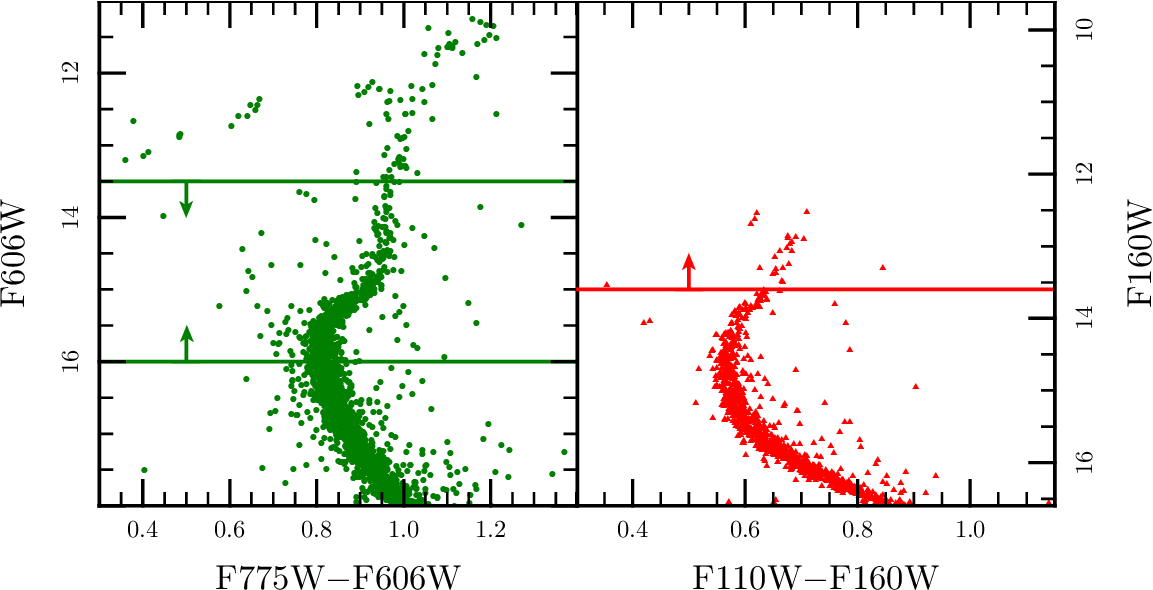}
\medskip

\includegraphics[width=\linewidth]{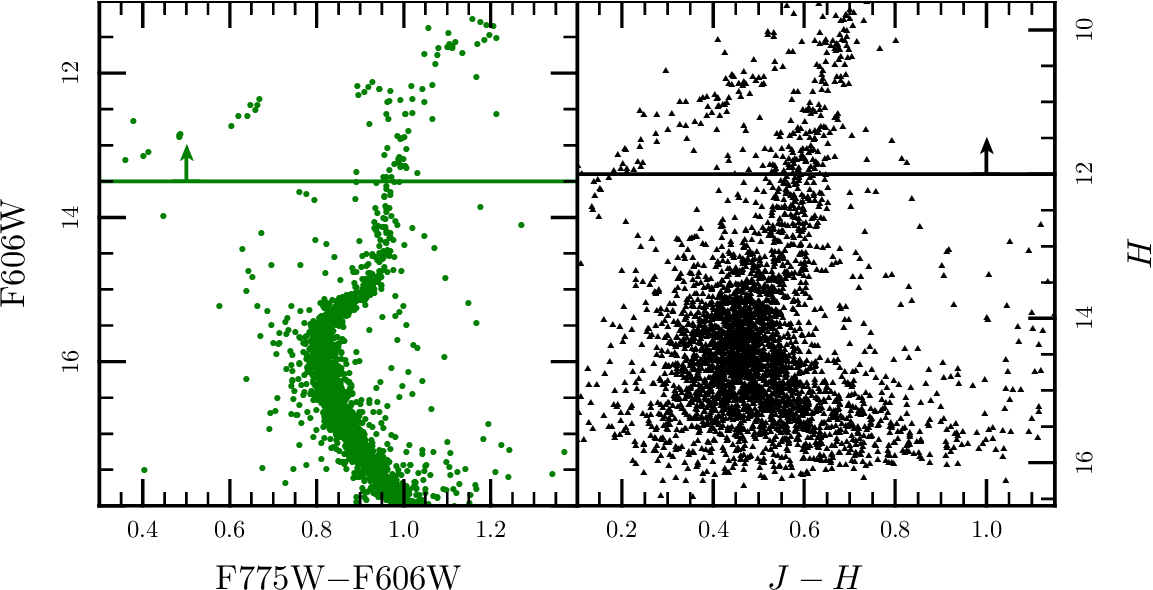}
\caption{The colour magnitude diagram for three overlapping samples in
  the globular cluster M4.  In the upper panels, the green points are
  a visual sample, and the red points are a IR sample.  The portion of
  the visual sample used to determine the coordinate transformation
  lies between the two green lines. The portion of the IR sample used lies
  above the red line.  The lower panels depict the overlapping samples
  used to connect the observations in the visual bands (left) to the
  2MASS observations (right).}
\label{fig:M4both}
\end{figure}

The $k$-d match algorithm is used to find the correspondances between
two stellar catalogues generated by Hubble Space Telescope (HST) data
and to connect these catalogues to the Two Micron All Sky Survey
(2MASS) catalogue, yielding absolute astrometry in the 2MASS system
\citep{2006AJ....131.1163S}.  \citet{2009ApJ...697..965B} observed a
field in the globular cluster Messier 4 to probe the faint end of the
white-dwarf cooling sequence and obtained measurements in the HST
bands F606W and F814W using the Advanced Camera for Survey (ACS).
Additional data in these bands were obtained as part of the program
GO-9578 (PI: Rhodes).  The second set of data is new from recent HST
observations by Dieball and collaborators in F110W and F160W using
the Wide-Field Camera 3 (WFC3).  The upper panel Fig.~\ref{fig:M4both}
depicts the colour-magnitude diagram of the two catalogues.  The
catalogue of objects detected in the visual bands is deeper, spans a
larger area and also contains brighter stars than the IR catalogue, so
taking the brighest thirty stars in each catalogue failed to find
matches; therefore, to find the possible correspondences, the IR
colours and magnitudes are plotted adjacent to those of the visual
bands to line up the turn-off in the two samples. The brightest 25
stars from the IR sample are matched against the 730 stars in the
visual sample whose magnitudes lie between 13.5 and 16.  There are
2,300 triangles in the IR sample and 64,569,960 in the visual sample.
No effort is made to cull triangles from the either sample.  It is
most efficient to build the $k$-d tree from the smaller sample; this
is what the package does by default.  A direct comparison of all the
triangles in the two catalogues numbers 148,510,908,000 and required
about three hours on a laptop computuer.  The $k$-d tree required
forty seconds on the same computer.  Of course, one could spend some
time to cull the second list of stars to perhaps thirty or so and
perform the direct comparison but this would have to be done carefully
because the overlap of the two fields is partial and would
probably require more than thirty seconds work.

\begin{figure}
\includegraphics[width=\linewidth]{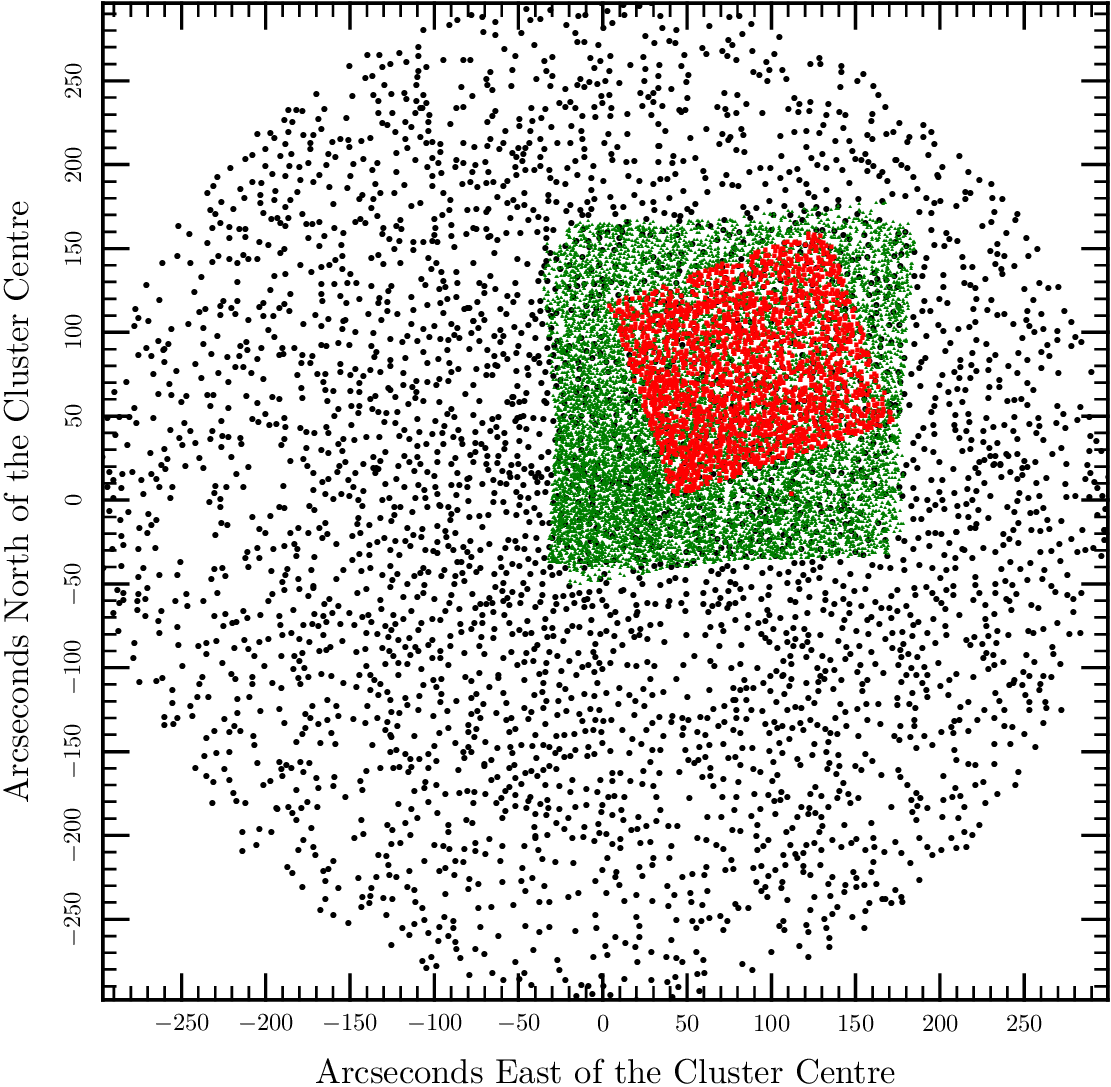}
\caption{The positions of the stars in the three overlapping samples in
  the globular cluster M4.  The green points are the visual sample, 
  the red points are the IR sample, and black points are point sources
  in the 2MASS catalogue.}
\label{fig:bothallplot}
\end{figure}

The second step is to find the transformation between the visual
sample and the 2MASS coordinate system using objects found in the
2MASS Point Source catalogue within five arcminutes of the centre of
M4 as depicted in the lower right panel of Fig.~\ref{fig:M4both}.
There are 88 stars in the visual sample and 354 in the 2MASS PSC.
Finding the best matching transformation between the two coordinate
systems requires about six seconds.  Both of these calculations show
the power of the $k$-d match algorithm.  In both cases a large
fraction of the stars in each catalogue are unique.  They don't appear
in both catalogues.  Finding the matching stars requires a more
exhaustive search than simply comparing say the brightest thirty stars
in each catalogue.  Although DAOMATCH \citep{1995PASP..107.1119V} was
not designed for matching catalogues with small fractional overlap,
DAOMATCH was used to try to find the best transformations between
these catalogues to provide a benchmark against a standard routine.
Because of the lack of overlap, DAOMATCH failed to find the
correspondences and generally stopped after comparing the first thirty
stars in each catalogue.  This exhaustive search typically takes about
0.5 seconds.  A similar exhaustive search with $k-$d match takes 0.007
seconds, about a factor of one hundred faster.

\begin{figure}
\includegraphics[width=\linewidth]{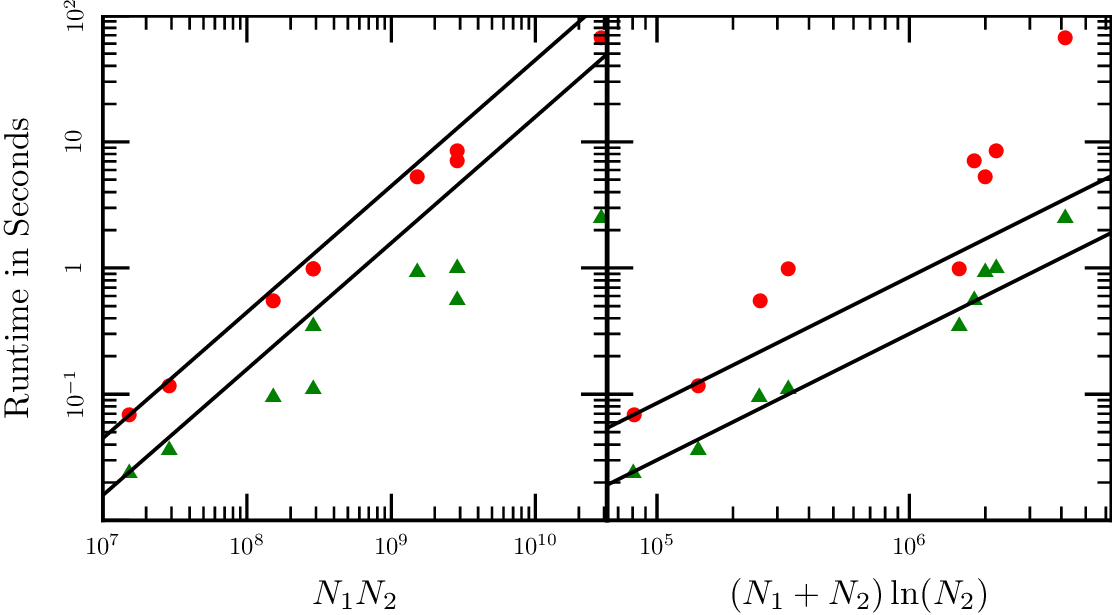}
\caption{The total runtime to match the objects in the IR catalogue to
those in the visual catalogue as a function of the number of objects
in each catalogue.  The red circles trace the results for the direct
technique, and the green triangles give the $k$-d tree algorithm.}
\label{fig:timeall}
\end{figure}

The triangle matching determines the coordinate transformations among
the IR sample, the visual sample and the 2MASS catalogue.
Fig.~\ref{fig:bothallplot} depicts the locations of the stars in the
IR sample in the coordinate system of the 2MASS Point Source
Catalogue.  The total IR sample contains 2,192 stars, the deeper and
wider visual sample contains 13,122 stars and the 2MASS PSC sample
contains 3761 stars.  The second step to merge the catagoues is also
dramatically accelerated.  To test the $k$-d match algorithm against a
variety of catalogue sizes, each catalogue is copied onto itself ten
or one hundred times and the nearest object in the second catalogue is
determined for each object in the first catalogue.  Other criteria are
possible.  For example, each object in the first catalogue can be
associated with all the objects within a given search radius in the
second catalogue.  Fig.~\ref{fig:timeall} depicts the runtime for each
of the algorithms as a function of the number of objects in each
catalogue.  The left panel gives the runtime as a function of the
product of the number of the objects in each catalogue, and the
runtime of the direct method is approximately proportional to this
product.  The right panel gives plots the runtime against the expected
logarithmic dependence of runtime for the $k$-d tree algorithm.  In
both cases the straight lines give a linear relation between the two
axes, normalized by the runtime of the smallest pair of catalogues.
Even for relatively small catalogues of a thousand objects, the $k$-d
tree algorithm outperforms the direct method.
\citet{Bentley:1975:MBS:361002.361007} argue the $k$-d tree search
should perform better than a direct search as along as there are more
than $2^k$ (i.e. four) nodes in the tree.

The second test uses same stars as the test for the triangle
algorithm.  In particular a random affine transformation is applied to
the coordinates of the brightest 25 stars of the IR sample, and this
transformed set of coordinates is matched against the original
coordinates.  The triangle algorithm fails to find any matching
triangles among the 2,300 asterisms and requires about 5~ms to run.
The quadrilateral matching technique (25,300 asterisms) takes
significantly longer, about 50~ms but consistently finds the correct
transformation.

\section{Discussion}
\label{sec:discussion}

This paper has outlined a new efficient algorithm for matching stellar
catalogues or lists of coordinates in general.  For large lists, the
speed-up relative to the generally available techniques that perform a
naive direct comparison are dramatic.  This speed-up allows the use of
larger source lists to accommodate catalogues in vastly different
bands and where one expects only a partial overlap between the lists
of objects.  The efficiency of the algorithm yields an efficient
implementation of the more difficult problem where the two coordinate
systems differ by an arbitrary affine transformation; that is when
shearing is present.  This may find application for astronomical
catalogues where the instrumentational shearing has not been corrected
or is unknown.  Additionally, this quadrilateral match algorithm could
also be applied more generally.  For example, if one ignores
perspective the rotation and projection of markers on or within
three-dimensional object into a two-dimensional image is equivalent to
a general affine transformation with a only few caveats when the
object is transparent.  Therefore, the quadrilateral matching
algorithm could be useful for applications as diverse as facial
recognition and the registration of medical and satellite imagery.

All of the routines outlined in this paper are available under the
``New BSD'' license at Sourceforge
(\url{https://sourceforge.net/projects/kdmatch}).  The reader is
encouraged to experiment with his or her data.

\section*{Acknowledgments}
JSH would like to thank Jason Kalirai for providing the star lists for
the bands F606W and F814W, Andrea Dieball for the star lists for F110W
and F160W, Stefan Reinsberg for useful discussions and the referee for
useful comments.  The research discussed is based on NASA/ESA Hubble
Space Telescope observations obtained at the Space Telescope Science
Institute, which is operated by the Association of Universities for
Research in Astronomy Inc. under NASA contract NAS5-26555. These
observations are associated with proposals GO-9578 (PI: Rhodes),
GO-10146 (PI: Bedin) and GO-12602 (PI: Dieball).  This publication
makes use of data products from the Two Micron All Sky Survey, which
is a joint project of the University of Massachusetts and the Infrared
Processing and Analysis Center/California Institute of Technology,
funded by the National Aeronautics and Space Administration and the
National Science Foundation.  This work was supported by the Natural
Sciences and Engineering Research Council of Canada.  It has made used
of the NASA ADS, arXiv.org, the Mikulski Archive for Space Telescopes
(MAST) and NASA/IPAC Infrared Science Archive (IRSA).

\bibliography{match}
\bibliographystyle{mn2e}

\label{lastpage}

\end{document}